\documentclass[aps, twocolumn, prx, longbibliography]{revtex4-1}
\usepackage{mathptmx} % use Times font instead of Computer Modern
\usepackage{amsfonts}
\usepackage{amsmath}
\usepackage{amssymb}
\usepackage{graphicx}
\usepackage{bm}
\usepackage{natbib}
\usepackage{color}
\usepackage[colorlinks=true,linkcolor=blue,anchorcolor=red,citecolor=blue, urlcolor=blue]{hyperref}
\usepackage{comment}

\begin{document}
\title{Energy gap tuning and gate-controlled topological phase transition in InAs/In$_{x}$Ga$_{1-x}$Sb composite quantum wells}
\author{H.~Irie}\email{hiroshi.irie.ke@hco.ntt.co.jp}\thanks{These authors contributed equally to this work.}
\author{T.~Akiho}\thanks{These authors contributed equally to this work.}
\author{F.~Cou\"{e}do}\altaffiliation[Present address: ]{Laboratoire National de Metrologie et d'Essais (LNE) Quantum Electrical Metrology Department, Avenue Roger Hennequin, 78197 Trappes, France}
\author{K.~Suzuki}\altaffiliation[Present address: ]{Fukuoka Institute of Technology, Fukuoka 811-0295, Japan}
\author{K.~Onomitsu}
\author{K.~Muraki}
\affiliation{NTT Basic Research Laboratories, NTT Corporation, 3-1 Morinosato-Wakamiya, Atsugi 243-0198, Japan}
\pacs{PACS number}
\date{\today}

\begin{abstract}
We report transport measurements of strained InAs/In$_{x}$Ga$_{1-x}$Sb composite quantum wells (CQWs) in the quantum spin Hall phase, focusing on the control of the energy gap through structural parameters and an external electric field. For highly strained CQWs with $x = 0.4$, we obtain a gap of 35 meV, an order of magnitude larger than that reported for binary InAs/GaSb CQWs. Using a dual-gate configuration, we demonstrate an electrical-field-driven topological phase transition, which manifests itself as a re-entrant behavior of the energy gap. The sizeable energy gap and high bulk resistivity obtained in both the topological and normal phases of a single device open the possibility of electrical switching of the edge transport.
\end{abstract}
\maketitle

Quantum spin Hall insulators (QSHIs), also known as two-dimensional time-reversal invariant topological insulators, have drawn increasing attention due to their spin-momentum-locked edge states, which are useful for spintronics applications and fault-tolerant quantum computation~\cite{Hasan2010, Qi2011, Kane2005, Bernevig2006science, Bernevig2006PRL, Konig2007}.
Since the edge states originate from the topologically nontrivial band structure of the bulk, their electronic properties are closely tied to those of the bulk state. Among various systems hosting a QSHI phase, InAs/GaSb composite quantum wells (CQWs) have a unique feature that the bulk properties (and concomitant edge properties) are tunable via non-material-specific parameters such as layer thickness and an electric field~\cite{Knez2010, Knez2011, Suzuki2013, Qu2015, Liu2008}.
The latter is brought about by the fact that the band inversion responsible for the QSHI phase arises not from the respective materials but from their relative band alignment~\cite{Liu2008}. Previous studies have demonstrated that the degree of band inversion $E_\mathrm{g0}$ varies with electric field~\cite{Knez2010, Qu2015, Suzuki2015, Shojaei2018} as well as with QW thickness~\cite{Suzuki2013, Yang1997, Jiang2017, Nichele2016}.
In particular, gate control of the band inversion across the phase transition at $E_\mathrm{g0} = 0$ allows one to switch the topological property \textit{in situ}, which adds a useful functionality for applications~\cite{Qu2015, Liu2008}.

While offering this \textit{in situ} tunability, the spatial separation of electron and hole wave functions into InAs and GaSb layers decreases their overlap and hence the energy gap $\Delta$ arising from the hybridization of electron and hole bands.
The resultant energy gap is at most 3 meV~\cite{Cooper1998, Yang1999}, which causes poor bulk insulation~\cite{Knez2010, Cooper1998, Charpentier2013} and inherently limits high-temperature operation.
As a solution to this problem, strain-engineered InAs/In$_{x}$Ga$_{1-x}$Sb CQWs have been proposed recently~\cite{Akiho2016, Du2017PRL}.
The in-plane biaxial compressive strain exerted on the In$_{x}$Ga$_{1-x}$Sb layer modifies the band structure, thus opening the possibility of achieving sizeable $\Delta$ while retaining the \textit{in situ} tunability.

In this paper, we report the control of the energy gap $\Delta$ in InAs/In$_{x}$Ga$_{1-x}$Sb CQWs through the InAs layer thickness $d_\mathrm{InAs}$ and alloy composition $x$ and demonstrate its \textit{in situ} tuning through gate voltages.
For highly strained CQWs with $x = 0.4$, we obtained $\Delta = 35$ meV (406 K), which is an order of magnitude larger than that reported for the binary InAs/GaSb system and is the largest in the ternary InAs/InGaSb system.
For a CQW with thickness close to the threshold of the band inversion, we demonstrate an electric-field-driven topological phase transition using a dual-gate configuration.
The phase transition manifests as a re-entrant behavior of the energy gap, with the gap increasing to sizable values ($> 10$ meV) on both the topological and normal insulator sides. We discuss several transport mechanisms expected at the transition point.

\begin{figure}[ptb]
\includegraphics{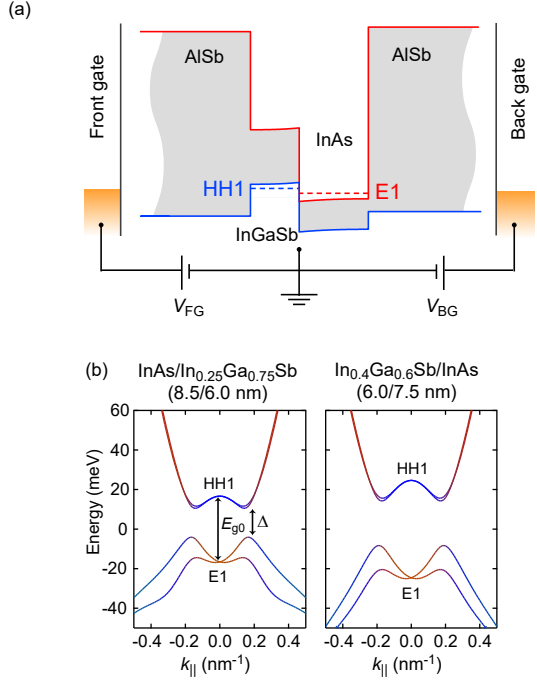}
\caption{\label{Fig1}(color online) (a) Band diagram of InAs/In$_{x}$Ga$_{1-x}$Sb CQW with AlSb barriers. (b) Band dispersions of HH1 and E1 subbands calculated for samples A (left) and B (right), which have $x = 0.2$ and 0.4 and $d_\mathrm{InAs} = 8.5$ and 7.5 nm, respectively. The color of the lines denotes the orbital character; i.e., red and blue represent electron-like and heavy-hole-like characters, respectively.}
\end{figure}

Figure~\ref{Fig1}(a) shows the band-edge profiles of InAs/In$_{x}$Ga$_{1-x}$Sb CQWs. Owing to the type-II broken-gap band alignment, electrons and holes are confined separately in the InAs and In$_{x}$Ga$_{1-x}$Sb layers, respectively.
For InAs and In$_{x}$Ga$_{1-x}$Sb layer thicknesses above certain critical values, the system has an inverted band structure, with the first electron subband (E1) located below the first heavy-hole subband (HH1). The degree of band inversion, quantified by $E_\mathrm{g0} \equiv E_\mathrm{HH1} - E_\mathrm{E1}$, where $E_\mathrm{E1}$ ($E_\mathrm{HH1}$) is the energy of the E1 (HH1) subband at the $\Gamma$ point, depends on the In composition $x$ of the In$_{x}$Ga$_{1-x}$Sb layer as well as the layer thicknesses.
The spatial separation of electron and hole wave functions allows $E_\mathrm{g0}$ to be controlled \textit{in situ} through an electric field applied from the front and back gates.

We studied molecular-beam-epitaxy-grown InAs/In$_{x}$Ga$_{1-x}$Sb/AlSb CQWs with $x = 0.25$ or 0.40 and various $d_\mathrm{InAs}$.
The first series of CQWs with $x = 0.25$, used in our previous study (Ref.~\cite{Akiho2016}), were grown on a Si-doped (001) GaAs substrate, while the second series with $x = 0.4$ were grown on a Te-doped (001) GaSb substrate.
In both cases, an 800-nm-thick AlSb buffer layer was employed, which resulted in full and partial ($\sim 50$\%) strain relaxation of the buffer layer on GaAs and GaSb substrates, respectively, as estimated by X-ray diffraction. For the $x = 0.25$ (0.4) CQWs, the layer order is such that InAs is above (below) In$_{x}$Ga$_{1-x}$Sb.
This layer order does not play any important roles in our results except for the dual-gate operation discussed later.
The thickness of the In$_{x}$Ga$_{1-x}$Sb layer $d_\mathrm{InGaSb}$ was fixed at 5.9 (6.0) nm for $x = 0.25$ (0.4), while $d_\mathrm{InAs}$ was varied as 8.5--11.9 (5.5--7.5) nm.
The CQW has a 50-nm-thick upper AlSb barrier, capped with a 5-nm-thick GaSb layer.
The samples were processed into 50-$\mu$m-wide Hall-bar devices (voltage probe distance of 180~$\mu$m) with a front gate fabricated on a 40-nm-thick atomic-layer-deposited aluminum oxide insulator.
The \textit{n}-type substrate was used as a back gate, which was grounded unless otherwise noted.
No hysteresis was observed within the gate-voltage range used.
All the samples were in the band-inverted regime except one with the smallest $d_\mathrm{InAs}$ of 5.5 nm (sample C, with $x = 0.4$), which was in the normal insulator phase at zero gate voltages.

Figure~\ref{Fig1}(b) shows the band dispersions for two representative samples, A and B (with $x = 0.25$ and 0.40 and $d_\mathrm{InAs} = 8.5$ and 7.5 nm, respectively), which were obtained by self-consistent eight-band $\mathbf{k\cdot p}$ calculations assuming zero gate voltages.
The red (blue) color shows the orbital character at each wave vector as being electron-like (heavy-hole-like), indicating that both samples have an inverted band structure (i.e., $E_\mathrm{g0} > 0$).
The biaxial compressive strain in the In$_{x}$Ga$_{1-x}$Sb layer increases the heavy-hole (HH)-light-hole (LH) splitting at the $\Gamma$ point, moving the HH and LH bands upwards and downward, respectively.
The former increases $E_\mathrm{g0}$ ($= E_\mathrm{HH1} - E_\mathrm{E1}$), which makes the band inversion achievable for thinner QW layers, allowing for an increased electron-hole wave function overlap.
The latter helps to suppress unwanted electron-LH mixing, which would otherwise complicate the band structure~\cite{Irie2020}.
[Note that the LH band is outside the range in Fig.~\ref{Fig1}(b); consequently, the LH-like orbital character represented by the green color is barely seen in Fig.~\ref{Fig1}(b).]
Both of these serve to enlarge the energy gap $\Delta$~\cite{Akiho2016, Du2017PRL} that opens at a finite in-plane momentum $k_\mathrm{cross}$ as a result of the hybridization between E1 and HH1 subbands.

\begin{figure}[ptb]
\includegraphics{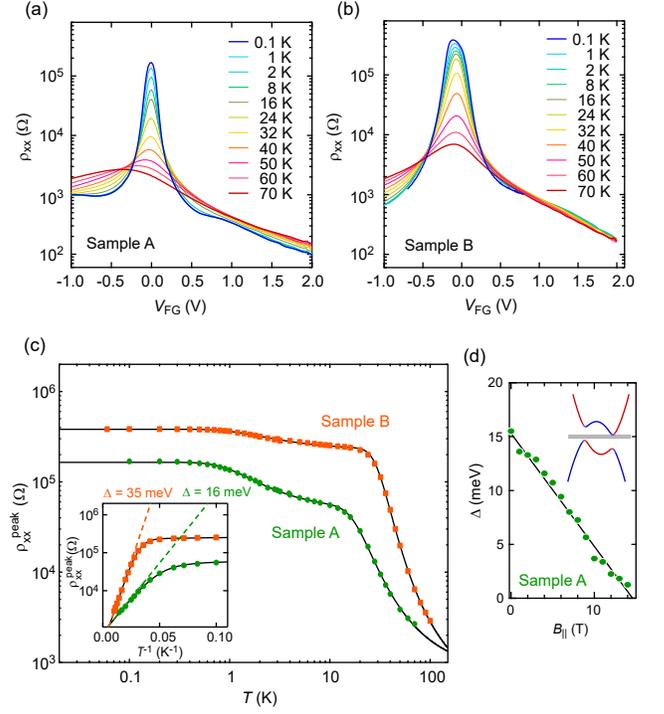}
\caption{\label{Fig2}(color online) (a), (b) $V_\mathrm{FG}$ dependence of $\mathrm{\rho}_{xx}$ at different temperatures for (a) sample A and (b) sample B. (c) Temperature dependence of the peak resistivity $\mathrm{\rho}_{xx}^{\mathrm{peak}}$. The solid lines are fits with a double exponential function. The inset is an Arrhenius plot of the data in the high-temperature regime. The solid (broken) lines are fits with a double (single) exponential function. (d) Variation of the energy gap $\Delta$ as a function of in-plane magnetic field $B_{||}$ (sample A). The inset illustrates the band dispersion under finite $B_{||}$.}
\end{figure}

We determined $\Delta$ for each sample from the temperature dependence of the longitudinal resistivity $\mathrm{\rho}_{xx}$.
Figure~\ref{Fig2}(a) shows $\mathrm{\rho}_{xx}$ of samples A ($x = 0.25$, $d_\mathrm{InAs} = 8.5$ nm), measured at various temperatures as a function of front-gate voltage $V_\mathrm{FG}$.
At low temperatures, $\mathrm{\rho}_{xx}$ displays a sharp peak at $V_\mathrm{FG} = -0.05$~V, which reflects the opening of an energy gap.
The peak is flanked by a dip on both sides, which is commonly seen in inverted InAs/(In)GaSb CQWs and has been attributed to the Van Hove singularity on the gap edge~\cite{Knez2010, Karalic2016}.
Fourier analysis of Shubnikov-de Haas oscillations (not shown) indicated the coexistence of electrons and holes, which confirmed that this sample has an inverted band structure~\cite{Akiho2016}.
At higher temperatures, the peak value of the resistivity ($\mathrm{\rho}_{xx}^{\mathrm{peak}}$) decreases with increasing temperature.
Figure~\ref{Fig2}(b) shows the result for sample B ($x = 0.40$, $d_\mathrm{InAs} = 7.5$ nm), where we observe similar $V_\mathrm{FG}$ and $T$ dependences, with a broader and higher $\mathrm{\rho}_{xx}$ peak at low temperatures.

Figure~\ref{Fig2}(c) summarizes the temperature dependence of $\mathrm{\rho}_{xx}^{\mathrm{peak}}$ for samples A and B.
For both samples, the observed $T$ dependence can be well described by a double exponential function,
\begin{equation}
  \label{Eq1}
  \left(\mathrm{\rho}_{xx}^{\mathrm{peak}}\right)^{-1} =
   G_{1}\exp\left(-\frac{\mathrm{\Delta}_{1}}{2k_{\mathrm{B}}T}\right) +
   G_{2}\exp\left(-\frac{\mathrm{\Delta}_{2}}{2k_{\mathrm{B}}T}\right) + G_{0},
\end{equation}
where $k_\mathrm{B}$ is the Boltzmann constant, and $G_{i}$ ($i =$ 0, 1, 2) and $\mathrm{\Delta}_{i}$ ($i =$ 1, 2) are constants with the dimensions of conductivity and energy, respectively.
The solid lines in Fig.~\ref{Fig2}(c) represent the results of the fitting, which yielded $\mathrm{\Delta}_{1} = 16$ meV and $\mathrm{\Delta}_{2} = 0.4$ meV for sample A and $\mathrm{\Delta}_{1} = 35$ meV and $\mathrm{\Delta}_{2} = 0.4$ meV for sample B.
(We define $\mathrm{\Delta}_{1} > \mathrm{\Delta}_{2}$.)

The first term of Eq. (\ref{Eq1}) is responsible for the $T$ dependence in the high-$T$ regime ($T > \sim$20 K).
The Arrhenius plot of the data in the high-$T$ regime shows that $\mathrm{\rho}_{xx}^{\mathrm{peak}}$ follows the activated $T$ dependence [the inset of Fig.~\ref{Fig2}(c)].
We thus identify $\mathrm{\Delta}_{1}$ as the size of the hybridization gap $\Delta$
\footnote{
Conventional Arrhenius analysis (i.e., fitting with a single exponential function) using high-$T$ data (40 K $\le T \le 70$ K) underestimates $\Delta$ by about 8\% compared to $\mathrm{\Delta}_{1}$.
}.
This is corroborated by measuring $\mathrm{\Delta}_{1}$ under an in-plane magnetic field $B_{||}$.
Figure~\ref{Fig2}(d) plots $\mathrm{\Delta}_{1}$ of sample A as a function of $B_{||}$.
The linear suppression of $\mathrm{\Delta}_{1}$ with $B_{||}$ is a manifestation of the gap shrinkage due to the relative shift of electron and hole bands in momentum space~\cite{Qu2015}, as illustrated in the inset of Fig.~\ref{Fig2}(d).

Below 20 K, the temperature dependence of $\mathrm{\rho}_{xx}^{\mathrm{peak}}$ becomes weak and then saturates, as described by the second and third terms of Eq.~(\ref{Eq1}).
Although such behavior has been commonly observed for inverted InAs/(In)GaSb CQWs~\cite{Knez2010, Suzuki2013, Suzuki2015, Du2015}, its origin is not fully understood.
Possibilities include the formation of a disorder-induced localization gap~\cite{Du2015} or the presence of electron-hole puddles arising from the spatial potential fluctuations~\cite{Suzuki2013}.
Although separate measurements using the Corbino geometry gave a slightly higher resistivity (not shown), suggesting a finite edge contribution in Hall-bar devices, similar low-temperature saturation was seen also in the Corbino devices.
While more study is needed to clarify and eliminate the cause of the residual conduction in the low-$T$ regime, we emphasize that it does not affect the following analyses of $\Delta$ using high-$T$ data.

\begin{figure}[ptb]
\includegraphics{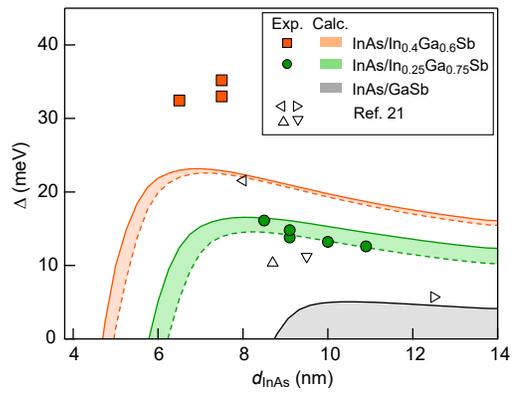}
\caption{\label{Fig3}(color online) Measured and calculated energy gap $\Delta$ of InAs/In$_{x}$Ga$_{1-x}$Sb CQWs in the inverted regime as a function of $d_\mathrm{InAs}$. The symbols represent measured $\Delta$.
Data reported in Ref.~\cite{Du2017PRL} for different $x$ and $d_\mathrm{InGaSb}$ are also plotted for comparison; $\triangleleft$, $x = 0.32$ and $d_\mathrm{InGaSb} = 4$ nm; $\triangle$, $x = 0.2$ and $d_\mathrm{InGaSb} = 4$ nm; $\triangledown$, $x = 0.25$ and $d_\mathrm{InGaSb} = 4$ nm; $\triangleright$, $x = 0$ and $d_\mathrm{GaSb} = 10$ nm.
The solid and broken lines, enclosing the shaded areas, represent $\Delta$ obtained from the self-consistent 8-band $\mathbf{k\cdot p}$ calculation assuming that the CQWs are pseudomorphically grown on a fully relaxed GaSb and AlSb layer, respectively.}
\end{figure}

We carried out similar measurements and analyses for CQWs with different $x$ and $d_\mathrm{InAs}$.
The obtained $\Delta$ values are plotted in Fig.~\ref{Fig3} as a function of $d_\mathrm{InAs}$ and compared with previously reported data~\cite{Du2017PRL} and our self-consistent $\mathbf{k\cdot p}$ calculations.
The solid and broken lines show $\Delta$ in the inverted regime calculated assuming pseudomorphic growth of CQWs on unstrained GaSb and fully relaxed AlSb buffer layers, respectively
\footnote{
A similar calculation without self-consistent potential was reported in Ref.~\cite{Akiho2016}, where the effect of strain in the InAs layer was mistakenly underestimated. Since these two effects tend to counteract each other, including them both produces overall similar results.
}.
The shaded area between the two lines indicates the range over which $\Delta$ can vary with epitaxial strain.
As a reference, we also included the calculation for $x = 0$, for which only the solid line is shown.
This is because, for the pseudomorphic growth on AlSb, the biaxial tensile strain in the GaSb layer makes the system semimetallic (and hence $\Delta =$ 0) for $d_\mathrm{GaSb} > 5.2$ nm~\cite{Irie2020}.
The calculations predict a qualitatively similar $d_\mathrm{InAs}$ dependence for all $x$; when $d_\mathrm{InAs}$ reaches a critical value and the system enters the inverted regime, $\Delta$ first sharply rises and then declines slowly towards the deep inverted regime.

Our data for $x = 0.25$ agree well with the calculation.
The $\Delta$ values reported in Ref.~\cite{Du2017PRL} are also grossly consistent with the expected behavior, although $x$ and $d_\mathrm{InGaSb}$ differ from those used in the calculations.
For $x = 0.4$, we obtained a large gap of 32–-35 meV, with the largest value of 35 meV corresponding to an order of magnitude enhancement compared to that of InAs/GaSb CQWs (typically $< 3$ meV~\cite{Cooper1998, Yang1999}).
Notably, the measured $\Delta$ values for $x = 0.4$ are considerably greater than the calculated ones.
The reason for this discrepancy is unknown. Tuning the material and structural parameters, including the band offset and layer thicknesses, did not reproduce the large values of $\Delta$.
As we see below, the calculation also slightly underestimates the critical thickness for the topological phase transition at which $\Delta$ goes to zero.

\begin{figure}[ptb]
\includegraphics{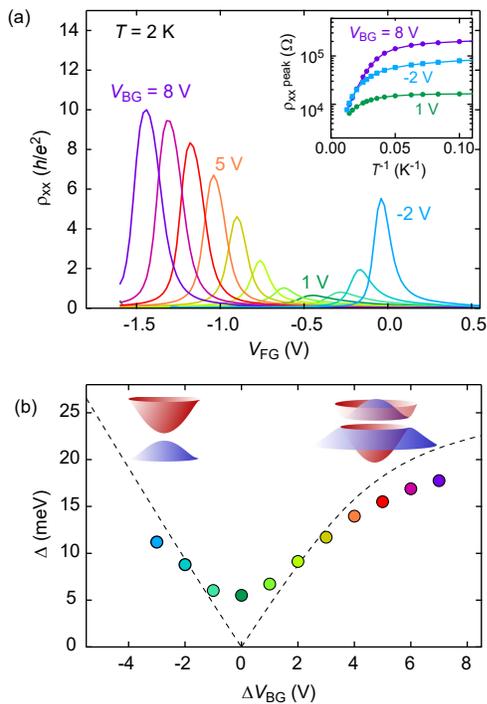}
\caption{\label{Fig4}(color online) (a) $\mathrm{\rho}_{xx}$ as a function of $V_\mathrm{FG}$ for sample C ($x = 0.4$, $d_\mathrm{InAs} = 5.5$ nm), measured at various $V_\mathrm{BG}$'s. The inset shows an Arrhenius plot for representative $V_\mathrm{BG}$ values. (b) $\Delta$ as a function of $\Delta V_\mathrm{BG}$, the shift in $V_\mathrm{BG}$ from the topological transition at $V_\mathrm{BG} = 1$ V.
$\Delta V_\mathrm{BG} > 0$ ($\Delta V_\mathrm{BG} < 0$) corresponds to the topological (normal) insulator phase. The broken line represents $\Delta$ obtained from the $\mathbf{k\cdot p}$ calculation. Note that $\Delta$ for $\Delta V_\mathrm{BG} < 0$ represents the energy gap in the normal insulator phase.}
\end{figure}

Having clarified how $\Delta$ depends on the structural parameters, now we examine the \textit{in situ} electric-field tuning of $\Delta$ and demonstrate a gate-controlled topological phase transition.
For this purpose, we use sample C ($x = 0.4$, $d_\mathrm{InAs} = 5.5$ nm), which was designed so that $E_\mathrm{g0}$ is close to zero at zero gate voltages.
As the InAs layer resides below the InGaSb layer in this sample, a positive (negative) back-gate voltage $V_\mathrm{BG}$ increases (decreases) $E_\mathrm{g0}$ and makes the system more (less) inverted.
Figure~\ref{Fig4}(a) shows $\mathrm{\rho}_{xx}$ vs.~$V_\mathrm{FG}$ measured at various $V_\mathrm{BG}$'s. With increasing $V_\mathrm{BG}$, the $\mathrm{\rho}_{xx}$ peak shifts to a more negative $V_\mathrm{FG}$, reflecting the corresponding charge neutrality condition.
The important observation is that the $\mathrm{\rho}_{xx}^{\mathrm{peak}}$ first decreases and then starts growing.
This re-entrant behavior becomes more evident by plotting $\Delta$ measured at each $V_\mathrm{BG}$ [Fig.~\ref{Fig4}(b)]
\footnote{
We used an Arrhenius analysis of high-$T$ data (30 K $\le T \le 60$ K) to obtain the $\Delta$ shown in Fig.~\ref{Fig4}(b), as fitting the weak temperature dependence near the topological phase transition with Eq.~(\ref{Eq1}) did not give reliable results.
},
demonstrating a gate-controlled topological-to-normal insulator phase transition.
The observed re-entrant $V_\mathrm{BG}$ dependence of $\Delta$ agrees fairly well with our self-consistent $\mathbf{k\cdot p}$ calculation, shown by the broken line in Fig.~\ref{Fig4}(b).

The significance of our data lies in the large $\Delta$ and, accordingly, the large $\mathrm{\rho}_{xx}^{\mathrm{peak}}$ achieved in both normal and topological insulator regimes.
Although a gate-controlled topological phase transition has previously been reported for InAs/GaSb CQWs~\cite{Qu2015, Shojaei2018}, the behavior of the gap was not examined therein.
Indeed, the data reported in Refs.~\cite{Qu2015, Shojaei2018} indicated significant residual bulk conduction, especially in the topological-insulator regime.
The high bulk resistivity ($\sim 10$ $h/e^{2}$) of our InAs/In$_{0.4}$Ga$_{0.6}$Sb CQW achieved in the vicinity of the phase transition is particularly favorable for the proposed topological transistor operation based on the switching of edge transport with gate voltages~\cite{Vandenberghe2017}.
At present, the maximum $\Delta$ in the normal-insulator phase is limited by the lowest $V_\mathrm{BG}$ of $- 2$~V that can be applied without causing hysteresis.
By improving the quality of the AlSb buffer layer, the bulk insulation in the normal-insulator phase is expected to be further improved.

The observation of a gate-controlled topological phase transition allows us to investigate the behavior right at the transition point, where one expects that the bulk gap closes and the band dispersion becomes similar to that of massless Dirac fermions.
Contrary to this expectation, an Arrhenius analysis of the temperature dependence yields a finite energy gap, as shown in Fig.~\ref{Fig4}(b).
Here, several caveats need to be considered. The first is that the temperature dependence at the transition is small---$\mathrm{\rho}_{xx}^{\mathrm{peak}}$ changes only by a factor of two [inset of Fig.~\ref{Fig4}(a)], which can be fitted by other functional forms.
Indeed, different temperature dependence, either quadratic or linear, is expected when the gap closes, depending on the ratio between temperature and level broadening~\cite{Buttner2011}.
The second is the electrostatic potential fluctuations due to background charge, which causes the formation of \textit{n}- or \textit{p}-type domains—or charge puddles—containing electrons or holes.
There, charge transport occurs through hopping or tunneling between charge puddles~\cite{Mahmoodian2020}.
The third is the spatial fluctuations in the CQW layer thicknesses and alloy composition that are unavoidable in actual samples.
As seen in Fig.~\ref{Fig3}, even a small increase in $d_\mathrm{InAs}$ by one monolayer (0.3 nm) around the critical thickness yields a large gap of about 10 meV.
Importantly, thickness variation around the critical value causes $E_\mathrm{g0}$ to spatially fluctuate around zero, which would result in the formation of topological and non-topological domains.
In this situation, transport can occur through helical ``edge'' channels that form in the sample interior along the borders between topologically distinct domains.
Furthermore, the combination of electrostatic potential fluctuations and the layer thickness fluctuations yields various transport regimes~\cite{Mahmoodian2020}.
The last is the possibility of the excitonic phases that are expected to emerge near the phase transition point~\cite{Naveh1996, Pikulin2014, Du2017NatCom}.
In such phases, the system remains an insulator, with the gap arising from the attractive Coulomb interaction between electrons and holes that form excitons.
A further investigation is needed for a detailed understanding of the behavior at the transition.

In summary, we reported the tuning of the energy gap in InAs/In$_{x}$Ga$_{1-x}$Sb CQWs by means of epitaxial strain, layer thicknesses, and an electric field.
The largest energy gap of 35 meV obtained in highly strained InAs/In$_{0.4}$Ga$_{0.6}$Sb CQWs is an order of magnitude larger than that for conventional InAs/GaSb CQWs.
While this value is not far from those of strain-engineered HgTe/CdHgTe QWs (55 meV)~\cite{Leubner2016} and monolayer WTe$_2$ (55 meV)~\cite{Tang2017}, the bulk resistivity at low temperatures is limited by a different mechanism with a smaller energy scale, indicating the need for further material improvement.
The electric-field driven topological phase transition with a sizeable energy gap in both topological and normal insulator phases opens the possibility of electrical switching of the edge transport.

The authors thank H. Murofushi for sample processing and M. Kamiya, R. Ohana, and H. Parke for transport measurements. This work was supported by JSPS KAKENHI (No. JP15H05854 and No. JP26287068).

H.I. and T.A. contributed equally to this work.

%\bibliography{largegap.bib}

\begin{thebibliography}{35}%
\makeatletter
\providecommand \@ifxundefined [1]{%
 \@ifx{#1\undefined}
}%
\providecommand \@ifnum [1]{%
 \ifnum #1\expandafter \@firstoftwo
 \else \expandafter \@secondoftwo
 \fi
}%
\providecommand \@ifx [1]{%
 \ifx #1\expandafter \@firstoftwo
 \else \expandafter \@secondoftwo
 \fi
}%
\providecommand \natexlab [1]{#1}%
\providecommand \enquote  [1]{``#1''}%
\providecommand \bibnamefont  [1]{#1}%
\providecommand \bibfnamefont [1]{#1}%
\providecommand \citenamefont [1]{#1}%
\providecommand \href@noop [0]{\@secondoftwo}%
\providecommand \href [0]{\begingroup \@sanitize@url \@href}%
\providecommand \@href[1]{\@@startlink{#1}\@@href}%
\providecommand \@@href[1]{\endgroup#1\@@endlink}%
\providecommand \@sanitize@url [0]{\catcode `\\12\catcode `\$12\catcode
  `\&12\catcode `\#12\catcode `\^12\catcode `\_12\catcode `\%12\relax}%
\providecommand \@@startlink[1]{}%
\providecommand \@@endlink[0]{}%
\providecommand \url  [0]{\begingroup\@sanitize@url \@url }%
\providecommand \@url [1]{\endgroup\@href {#1}{\urlprefix }}%
\providecommand \urlprefix  [0]{URL }%
\providecommand \Eprint [0]{\href }%
\providecommand \doibase [0]{http://dx.doi.org/}%
\providecommand \selectlanguage [0]{\@gobble}%
\providecommand \bibinfo  [0]{\@secondoftwo}%
\providecommand \bibfield  [0]{\@secondoftwo}%
\providecommand \translation [1]{[#1]}%
\providecommand \BibitemOpen [0]{}%
\providecommand \bibitemStop [0]{}%
\providecommand \bibitemNoStop [0]{.\EOS\space}%
\providecommand \EOS [0]{\spacefactor3000\relax}%
\providecommand \BibitemShut  [1]{\csname bibitem#1\endcsname}%
\let\auto@bib@innerbib\@empty
%</preamble>
\bibitem [{\citenamefont {Hasan}\ and\ \citenamefont {Kane}(2010)}]{Hasan2010}%
  \BibitemOpen
  \bibfield  {author} {\bibinfo {author} {\bibfnamefont {M.~Z.}\ \bibnamefont
  {Hasan}}\ and\ \bibinfo {author} {\bibfnamefont {C.~L.}\ \bibnamefont
  {Kane}},\ }\bibfield  {title} {\enquote {\bibinfo {title} {{Colloquium:
  Topological insulators}},}\ }\href {\doibase 10.1103/RevModPhys.82.3045}
  {\bibfield  {journal} {\bibinfo  {journal} {Rev. Mod. Phys.}\ }\textbf
  {\bibinfo {volume} {82}},\ \bibinfo {pages} {3045} (\bibinfo {year}
  {2010})}\BibitemShut {NoStop}%
\bibitem [{\citenamefont {Qi}\ and\ \citenamefont {Zhang}(2011)}]{Qi2011}%
  \BibitemOpen
  \bibfield  {author} {\bibinfo {author} {\bibfnamefont {X.-L.}\ \bibnamefont
  {Qi}}\ and\ \bibinfo {author} {\bibfnamefont {S.-C.}\ \bibnamefont {Zhang}},\
  }\bibfield  {title} {\enquote {\bibinfo {title} {{Topological insulators and
  superconductors}},}\ }\href {\doibase 10.1103/RevModPhys.83.1057} {\bibfield
  {journal} {\bibinfo  {journal} {Rev. Mod. Phys.}\ }\textbf {\bibinfo {volume}
  {83}},\ \bibinfo {pages} {1057} (\bibinfo {year} {2011})}\BibitemShut
  {NoStop}%
\bibitem [{\citenamefont {Kane}\ and\ \citenamefont {Mele}(2005)}]{Kane2005}%
  \BibitemOpen
  \bibfield  {author} {\bibinfo {author} {\bibfnamefont {C.~L.}\ \bibnamefont
  {Kane}}\ and\ \bibinfo {author} {\bibfnamefont {E.~J.}\ \bibnamefont
  {Mele}},\ }\bibfield  {title} {\enquote {\bibinfo {title} {${Z}_{2}$
  topological order and the quantum spin Hall effect},}\ }\href {\doibase
  10.1103/PhysRevLett.95.146802} {\bibfield  {journal} {\bibinfo  {journal}
  {Phys. Rev. Lett.}\ }\textbf {\bibinfo {volume} {95}},\ \bibinfo {pages}
  {146802} (\bibinfo {year} {2005})}\BibitemShut {NoStop}%
\bibitem [{\citenamefont {Bernevig}\ \emph {et~al.}(2006)\citenamefont
  {Bernevig}, \citenamefont {Hughes},\ and\ \citenamefont
  {Zhang}}]{Bernevig2006science}%
  \BibitemOpen
  \bibfield  {author} {\bibinfo {author} {\bibfnamefont {B.~A.}\ \bibnamefont
  {Bernevig}}, \bibinfo {author} {\bibfnamefont {T.~L.}\ \bibnamefont
  {Hughes}}, \ and\ \bibinfo {author} {\bibfnamefont {S.-C.}\ \bibnamefont
  {Zhang}},\ }\bibfield  {title} {\enquote {\bibinfo {title} {{Quantum Spin
  Hall Effect and Topological Phase Transition in HgTe Quantum Wells}},}\
  }\href {\doibase 10.1126/science.1133734} {\bibfield  {journal} {\bibinfo
  {journal} {Science}\ }\textbf {\bibinfo {volume} {314}},\ \bibinfo {pages}
  {1757} (\bibinfo {year} {2006})}\BibitemShut {NoStop}%
\bibitem [{\citenamefont {Bernevig}\ and\ \citenamefont
  {Zhang}(2006)}]{Bernevig2006PRL}%
  \BibitemOpen
  \bibfield  {author} {\bibinfo {author} {\bibfnamefont {B.~A.}\ \bibnamefont
  {Bernevig}}\ and\ \bibinfo {author} {\bibfnamefont {S.-C.}\ \bibnamefont
  {Zhang}},\ }\bibfield  {title} {\enquote {\bibinfo {title} {{Quantum Spin
  Hall Effect}},}\ }\href {\doibase 10.1103/PhysRevLett.96.106802} {\bibfield
  {journal} {\bibinfo  {journal} {Phys. Rev. Lett.}\ }\textbf {\bibinfo
  {volume} {96}},\ \bibinfo {pages} {106802} (\bibinfo {year}
  {2006})}\BibitemShut {NoStop}%
\bibitem [{\citenamefont {K{\"o}nig}\ \emph {et~al.}(2007)\citenamefont
  {K{\"o}nig}, \citenamefont {Wiedmann}, \citenamefont {Br{\"u}ne},
  \citenamefont {Roth}, \citenamefont {Buhmann}, \citenamefont {Molenkamp},
  \citenamefont {Qi},\ and\ \citenamefont {Zhang}}]{Konig2007}%
  \BibitemOpen
  \bibfield  {author} {\bibinfo {author} {\bibfnamefont {M.}~\bibnamefont
  {K{\"o}nig}}, \bibinfo {author} {\bibfnamefont {S.}~\bibnamefont {Wiedmann}},
  \bibinfo {author} {\bibfnamefont {C.}~\bibnamefont {Br{\"u}ne}}, \bibinfo
  {author} {\bibfnamefont {A.}~\bibnamefont {Roth}}, \bibinfo {author}
  {\bibfnamefont {H.}~\bibnamefont {Buhmann}}, \bibinfo {author} {\bibfnamefont
  {L.~W.}\ \bibnamefont {Molenkamp}}, \bibinfo {author} {\bibfnamefont {X.-L.}\
  \bibnamefont {Qi}}, \ and\ \bibinfo {author} {\bibfnamefont {S.-C.}\
  \bibnamefont {Zhang}},\ }\bibfield  {title} {\enquote {\bibinfo {title}
  {{Quantum Spin Hall Insulator State in HgTe Quantum Wells}},}\ }\href
  {\doibase 10.1126/science.1148047} {\bibfield  {journal} {\bibinfo  {journal}
  {Science}\ }\textbf {\bibinfo {volume} {318}},\ \bibinfo {pages} {766}
  (\bibinfo {year} {2007})}\BibitemShut {NoStop}%
\bibitem [{\citenamefont {Knez}\ \emph {et~al.}(2010)\citenamefont {Knez},
  \citenamefont {Du},\ and\ \citenamefont {Sullivan}}]{Knez2010}%
  \BibitemOpen
  \bibfield  {author} {\bibinfo {author} {\bibfnamefont {I.}~\bibnamefont
  {Knez}}, \bibinfo {author} {\bibfnamefont {R.~R.}\ \bibnamefont {Du}}, \ and\
  \bibinfo {author} {\bibfnamefont {G.}~\bibnamefont {Sullivan}},\ }\bibfield
  {title} {\enquote {\bibinfo {title} {{Finite conductivity in mesoscopic Hall
  bars of inverted InAs/GaSb quantum wells}},}\ }\href {\doibase
  10.1103/PhysRevB.81.201301} {\bibfield  {journal} {\bibinfo  {journal} {Phys.
  Rev. B}\ }\textbf {\bibinfo {volume} {81}},\ \bibinfo {pages} {201301}
  (\bibinfo {year} {2010})}\BibitemShut {NoStop}%
\bibitem [{\citenamefont {Knez}\ \emph {et~al.}(2011)\citenamefont {Knez},
  \citenamefont {Du},\ and\ \citenamefont {Sullivan}}]{Knez2011}%
  \BibitemOpen
  \bibfield  {author} {\bibinfo {author} {\bibfnamefont {I.}~\bibnamefont
  {Knez}}, \bibinfo {author} {\bibfnamefont {R.-R.}\ \bibnamefont {Du}}, \ and\
  \bibinfo {author} {\bibfnamefont {G.}~\bibnamefont {Sullivan}},\ }\bibfield
  {title} {\enquote {\bibinfo {title} {{Evidence for Helical Edge Modes in
  Inverted InAs/GaSb Quantum Wells}},}\ }\href {\doibase
  10.1103/PhysRevLett.107.136603} {\bibfield  {journal} {\bibinfo  {journal}
  {Phys. Rev. Lett.}\ }\textbf {\bibinfo {volume} {107}},\ \bibinfo {pages}
  {136603} (\bibinfo {year} {2011})}\BibitemShut {NoStop}%
\bibitem [{\citenamefont {Suzuki}\ \emph {et~al.}(2013)\citenamefont {Suzuki},
  \citenamefont {Harada}, \citenamefont {Onomitsu},\ and\ \citenamefont
  {Muraki}}]{Suzuki2013}%
  \BibitemOpen
  \bibfield  {author} {\bibinfo {author} {\bibfnamefont {K.}~\bibnamefont
  {Suzuki}}, \bibinfo {author} {\bibfnamefont {Y.}~\bibnamefont {Harada}},
  \bibinfo {author} {\bibfnamefont {K.}~\bibnamefont {Onomitsu}}, \ and\
  \bibinfo {author} {\bibfnamefont {K.}~\bibnamefont {Muraki}},\ }\bibfield
  {title} {\enquote {\bibinfo {title} {{Edge channel transport in the InAs/GaSb
  topological insulating phase}},}\ }\href {\doibase
  10.1103/PhysRevB.87.235311} {\bibfield  {journal} {\bibinfo  {journal} {Phys.
  Rev. B}\ }\textbf {\bibinfo {volume} {87}},\ \bibinfo {pages} {235311}
  (\bibinfo {year} {2013})}\BibitemShut {NoStop}%
\bibitem [{\citenamefont {Qu}\ \emph {et~al.}(2015)\citenamefont {Qu},
  \citenamefont {Beukman}, \citenamefont {Nadj-Perge}, \citenamefont {Wimmer},
  \citenamefont {Nguyen}, \citenamefont {Yi}, \citenamefont {Thorp},
  \citenamefont {Sokolich}, \citenamefont {Kiselev}, \citenamefont {Manfra},
  \citenamefont {Marcus},\ and\ \citenamefont {Kouwenhoven}}]{Qu2015}%
  \BibitemOpen
  \bibfield  {author} {\bibinfo {author} {\bibfnamefont {F.}~\bibnamefont
  {Qu}}, \bibinfo {author} {\bibfnamefont {A.~J.~A.}\ \bibnamefont {Beukman}},
  \bibinfo {author} {\bibfnamefont {S.}~\bibnamefont {Nadj-Perge}}, \bibinfo
  {author} {\bibfnamefont {M.}~\bibnamefont {Wimmer}}, \bibinfo {author}
  {\bibfnamefont {B.-M.}\ \bibnamefont {Nguyen}}, \bibinfo {author}
  {\bibfnamefont {W.}~\bibnamefont {Yi}}, \bibinfo {author} {\bibfnamefont
  {J.}~\bibnamefont {Thorp}}, \bibinfo {author} {\bibfnamefont
  {M.}~\bibnamefont {Sokolich}}, \bibinfo {author} {\bibfnamefont {A.~A.}\
  \bibnamefont {Kiselev}}, \bibinfo {author} {\bibfnamefont {M.~J.}\
  \bibnamefont {Manfra}}, \bibinfo {author} {\bibfnamefont {C.~M.}\
  \bibnamefont {Marcus}}, \ and\ \bibinfo {author} {\bibfnamefont {L.~P.}\
  \bibnamefont {Kouwenhoven}},\ }\bibfield  {title} {\enquote {\bibinfo {title}
  {{Electric and Magnetic Tuning Between the Trivial and Topological Phases in
  InAs/GaSb Double Quantum Wells}},}\ }\href {\doibase
  10.1103/PhysRevLett.115.036803} {\bibfield  {journal} {\bibinfo  {journal}
  {Phys. Rev. Lett.}\ }\textbf {\bibinfo {volume} {115}},\ \bibinfo {pages}
  {036803} (\bibinfo {year} {2015})}\BibitemShut {NoStop}%
\bibitem [{\citenamefont {Liu}\ \emph {et~al.}(2008)\citenamefont {Liu},
  \citenamefont {Hughes}, \citenamefont {Qi}, \citenamefont {Wang},\ and\
  \citenamefont {Zhang}}]{Liu2008}%
  \BibitemOpen
  \bibfield  {author} {\bibinfo {author} {\bibfnamefont {C.}~\bibnamefont
  {Liu}}, \bibinfo {author} {\bibfnamefont {T.~L.}\ \bibnamefont {Hughes}},
  \bibinfo {author} {\bibfnamefont {X.-L.}\ \bibnamefont {Qi}}, \bibinfo
  {author} {\bibfnamefont {K.}~\bibnamefont {Wang}}, \ and\ \bibinfo {author}
  {\bibfnamefont {S.-C.}\ \bibnamefont {Zhang}},\ }\bibfield  {title} {\enquote
  {\bibinfo {title} {{Quantum Spin Hall Effect in Inverted Type-II
  Semiconductors}},}\ }\href {\doibase 10.1103/PhysRevLett.100.236601}
  {\bibfield  {journal} {\bibinfo  {journal} {Phys. Rev. Lett.}\ }\textbf
  {\bibinfo {volume} {100}},\ \bibinfo {pages} {236601} (\bibinfo {year}
  {2008})}\BibitemShut {NoStop}%
\bibitem [{\citenamefont {Suzuki}\ \emph {et~al.}(2015)\citenamefont {Suzuki},
  \citenamefont {Harada}, \citenamefont {Onomitsu},\ and\ \citenamefont
  {Muraki}}]{Suzuki2015}%
  \BibitemOpen
  \bibfield  {author} {\bibinfo {author} {\bibfnamefont {K.}~\bibnamefont
  {Suzuki}}, \bibinfo {author} {\bibfnamefont {Y.}~\bibnamefont {Harada}},
  \bibinfo {author} {\bibfnamefont {K.}~\bibnamefont {Onomitsu}}, \ and\
  \bibinfo {author} {\bibfnamefont {K.}~\bibnamefont {Muraki}},\ }\bibfield
  {title} {\enquote {\bibinfo {title} {{Gate-controlled semimetal-topological
  insulator transition in an InAs/GaSb heterostructure}},}\ }\href {\doibase
  10.1103/PhysRevB.91.245309} {\bibfield  {journal} {\bibinfo  {journal} {Phys.
  Rev. B}\ }\textbf {\bibinfo {volume} {91}},\ \bibinfo {pages} {245309}
  (\bibinfo {year} {2015})}\BibitemShut {NoStop}%
\bibitem [{\citenamefont {Shojaei}\ \emph {et~al.}(2018)\citenamefont
  {Shojaei}, \citenamefont {McFadden}, \citenamefont {Pendharkar},
  \citenamefont {Lee}, \citenamefont {Flatt\'e},\ and\ \citenamefont
  {Palmstr\o{}m}}]{Shojaei2018}%
  \BibitemOpen
  \bibfield  {author} {\bibinfo {author} {\bibfnamefont {B.}~\bibnamefont
  {Shojaei}}, \bibinfo {author} {\bibfnamefont {A.~P.}\ \bibnamefont
  {McFadden}}, \bibinfo {author} {\bibfnamefont {M.}~\bibnamefont
  {Pendharkar}}, \bibinfo {author} {\bibfnamefont {J.~S.}\ \bibnamefont {Lee}},
  \bibinfo {author} {\bibfnamefont {M.~E.}\ \bibnamefont {Flatt\'e}}, \ and\
  \bibinfo {author} {\bibfnamefont {C.~J.}\ \bibnamefont {Palmstr\o{}m}},\
  }\bibfield  {title} {\enquote {\bibinfo {title} {{Materials considerations
  for forming the topological insulator phase in InAs/GaSb
  heterostructures}},}\ }\href {\doibase 10.1103/PhysRevMaterials.2.064603}
  {\bibfield  {journal} {\bibinfo  {journal} {Phys. Rev. Mater.}\ }\textbf
  {\bibinfo {volume} {2}},\ \bibinfo {pages} {064603} (\bibinfo {year}
  {2018})}\BibitemShut {NoStop}%
\bibitem [{\citenamefont {Yang}\ \emph {et~al.}(1997)\citenamefont {Yang},
  \citenamefont {Yang}, \citenamefont {Bennett},\ and\ \citenamefont
  {Shanabrook}}]{Yang1997}%
  \BibitemOpen
  \bibfield  {author} {\bibinfo {author} {\bibfnamefont {M.~J.}\ \bibnamefont
  {Yang}}, \bibinfo {author} {\bibfnamefont {C.~H.}\ \bibnamefont {Yang}},
  \bibinfo {author} {\bibfnamefont {B.~R.}\ \bibnamefont {Bennett}}, \ and\
  \bibinfo {author} {\bibfnamefont {B.~V.}\ \bibnamefont {Shanabrook}},\
  }\bibfield  {title} {\enquote {\bibinfo {title} {{Evidence of a Hybridization
  Gap in ``Semimetallic'' InAs/GaSb Systems}},}\ }\href {\doibase
  10.1103/PhysRevLett.78.4613} {\bibfield  {journal} {\bibinfo  {journal}
  {Phys. Rev. Lett.}\ }\textbf {\bibinfo {volume} {78}},\ \bibinfo {pages}
  {4613} (\bibinfo {year} {1997})}\BibitemShut {NoStop}%
\bibitem [{\citenamefont {Jiang}\ \emph {et~al.}(2017)\citenamefont {Jiang},
  \citenamefont {Thapa}, \citenamefont {Sanders}, \citenamefont {Stanton},
  \citenamefont {Zhang}, \citenamefont {Kono}, \citenamefont {Lou},
  \citenamefont {Chang}, \citenamefont {Hawkins}, \citenamefont {Klem},
  \citenamefont {Pan}, \citenamefont {Smirnov},\ and\ \citenamefont
  {Jiang}}]{Jiang2017}%
  \BibitemOpen
  \bibfield  {author} {\bibinfo {author} {\bibfnamefont {Y.}~\bibnamefont
  {Jiang}}, \bibinfo {author} {\bibfnamefont {S.}~\bibnamefont {Thapa}},
  \bibinfo {author} {\bibfnamefont {G.~D.}\ \bibnamefont {Sanders}}, \bibinfo
  {author} {\bibfnamefont {C.~J.}\ \bibnamefont {Stanton}}, \bibinfo {author}
  {\bibfnamefont {Q.}~\bibnamefont {Zhang}}, \bibinfo {author} {\bibfnamefont
  {J.}~\bibnamefont {Kono}}, \bibinfo {author} {\bibfnamefont {W.~K.}\
  \bibnamefont {Lou}}, \bibinfo {author} {\bibfnamefont {K.}~\bibnamefont
  {Chang}}, \bibinfo {author} {\bibfnamefont {S.~D.}\ \bibnamefont {Hawkins}},
  \bibinfo {author} {\bibfnamefont {J.~F.}\ \bibnamefont {Klem}}, \bibinfo
  {author} {\bibfnamefont {W.}~\bibnamefont {Pan}}, \bibinfo {author}
  {\bibfnamefont {D.}~\bibnamefont {Smirnov}}, \ and\ \bibinfo {author}
  {\bibfnamefont {Z.}~\bibnamefont {Jiang}},\ }\bibfield  {title} {\enquote
  {\bibinfo {title} {{Probing the semiconductor to semimetal transition in
  InAs/GaSb double quantum wells by magneto-infrared spectroscopy}},}\ }\href
  {\doibase 10.1103/PhysRevB.95.045116} {\bibfield  {journal} {\bibinfo
  {journal} {Phys. Rev. B}\ }\textbf {\bibinfo {volume} {95}},\ \bibinfo
  {pages} {045116} (\bibinfo {year} {2017})}\BibitemShut {NoStop}%
\bibitem [{\citenamefont {Nichele}\ \emph {et~al.}(2016)\citenamefont
  {Nichele}, \citenamefont {Suominen}, \citenamefont {Kjaergaard},
  \citenamefont {Marcus}, \citenamefont {Sajadi}, \citenamefont {Folk},
  \citenamefont {Qu}, \citenamefont {Beukman}, \citenamefont {de~Vries},
  \citenamefont {van Veen}, \citenamefont {Nadj-Perge}, \citenamefont
  {Kouwenhoven}, \citenamefont {Nguyen}, \citenamefont {Kiselev}, \citenamefont
  {Yi}, \citenamefont {Sokolich}, \citenamefont {Manfra}, \citenamefont
  {Spanton},\ and\ \citenamefont {Moler}}]{Nichele2016}%
  \BibitemOpen
  \bibfield  {author} {\bibinfo {author} {\bibfnamefont {F.}~\bibnamefont
  {Nichele}}, \bibinfo {author} {\bibfnamefont {H.~J.}\ \bibnamefont
  {Suominen}}, \bibinfo {author} {\bibfnamefont {M.}~\bibnamefont
  {Kjaergaard}}, \bibinfo {author} {\bibfnamefont {C.~M.}\ \bibnamefont
  {Marcus}}, \bibinfo {author} {\bibfnamefont {E.}~\bibnamefont {Sajadi}},
  \bibinfo {author} {\bibfnamefont {J.~A.}\ \bibnamefont {Folk}}, \bibinfo
  {author} {\bibfnamefont {F.}~\bibnamefont {Qu}}, \bibinfo {author}
  {\bibfnamefont {A.~J.~A.}\ \bibnamefont {Beukman}}, \bibinfo {author}
  {\bibfnamefont {F.~K.}\ \bibnamefont {de~Vries}}, \bibinfo {author}
  {\bibfnamefont {J.}~\bibnamefont {van Veen}}, \bibinfo {author}
  {\bibfnamefont {S.}~\bibnamefont {Nadj-Perge}}, \bibinfo {author}
  {\bibfnamefont {L.~P.}\ \bibnamefont {Kouwenhoven}}, \bibinfo {author}
  {\bibfnamefont {B.~M.}\ \bibnamefont {Nguyen}}, \bibinfo {author}
  {\bibfnamefont {A.~A.}\ \bibnamefont {Kiselev}}, \bibinfo {author}
  {\bibfnamefont {W.}~\bibnamefont {Yi}}, \bibinfo {author} {\bibfnamefont
  {M.}~\bibnamefont {Sokolich}}, \bibinfo {author} {\bibfnamefont {M.~J.}\
  \bibnamefont {Manfra}}, \bibinfo {author} {\bibfnamefont {E.~M.}\
  \bibnamefont {Spanton}}, \ and\ \bibinfo {author} {\bibfnamefont {K.~A.}\
  \bibnamefont {Moler}},\ }\bibfield  {title} {\enquote {\bibinfo {title}
  {{Edge transport in the trivial phase of InAs/GaSb}},}\ }\href {\doibase
  10.1088/1367-2630/18/8/083005} {\bibfield  {journal} {\bibinfo  {journal}
  {New J. Phys.}\ }\textbf {\bibinfo {volume} {18}},\ \bibinfo {pages} {083005}
  (\bibinfo {year} {2016})}\BibitemShut {NoStop}%
\bibitem [{\citenamefont {Cooper}\ \emph {et~al.}(1998)\citenamefont {Cooper},
  \citenamefont {Patel}, \citenamefont {Drouot}, \citenamefont {Linfield},
  \citenamefont {Ritchie},\ and\ \citenamefont {Pepper}}]{Cooper1998}%
  \BibitemOpen
  \bibfield  {author} {\bibinfo {author} {\bibfnamefont {L.~J.}\ \bibnamefont
  {Cooper}}, \bibinfo {author} {\bibfnamefont {N.~K.}\ \bibnamefont {Patel}},
  \bibinfo {author} {\bibfnamefont {V.}~\bibnamefont {Drouot}}, \bibinfo
  {author} {\bibfnamefont {E.~H.}\ \bibnamefont {Linfield}}, \bibinfo {author}
  {\bibfnamefont {D.~A.}\ \bibnamefont {Ritchie}}, \ and\ \bibinfo {author}
  {\bibfnamefont {M.}~\bibnamefont {Pepper}},\ }\bibfield  {title} {\enquote
  {\bibinfo {title} {{Resistance resonance induced by electron-hole
  hybridization in a strongly coupled InAs/GaSb/AlSb heterostructure}},}\
  }\href {\doibase 10.1103/PhysRevB.57.11915} {\bibfield  {journal} {\bibinfo
  {journal} {Phys. Rev. B}\ }\textbf {\bibinfo {volume} {57}},\ \bibinfo
  {pages} {11915} (\bibinfo {year} {1998})}\BibitemShut {NoStop}%
\bibitem [{\citenamefont {Yang}\ \emph {et~al.}(1999)\citenamefont {Yang},
  \citenamefont {Yang},\ and\ \citenamefont {Bennett}}]{Yang1999}%
  \BibitemOpen
  \bibfield  {author} {\bibinfo {author} {\bibfnamefont {M.~J.}\ \bibnamefont
  {Yang}}, \bibinfo {author} {\bibfnamefont {C.~H.}\ \bibnamefont {Yang}}, \
  and\ \bibinfo {author} {\bibfnamefont {B.~R.}\ \bibnamefont {Bennett}},\
  }\bibfield  {title} {\enquote {\bibinfo {title} {{Magnetocapacitance and
  far-infrared photoconductivity in GaSb/InAs composite quantum wells}},}\
  }\href {\doibase 10.1103/PhysRevB.60.R13958} {\bibfield  {journal} {\bibinfo
  {journal} {Phys. Rev. B}\ }\textbf {\bibinfo {volume} {60}},\ \bibinfo
  {pages} {R13958} (\bibinfo {year} {1999})}\BibitemShut {NoStop}%
\bibitem [{\citenamefont {Charpentier}\ \emph {et~al.}(2013)\citenamefont
  {Charpentier}, \citenamefont {F{\"{a}}lt}, \citenamefont {Reichl},
  \citenamefont {Nichele}, \citenamefont {Nath~Pal}, \citenamefont {Pietsch},
  \citenamefont {Ihn}, \citenamefont {Ensslin},\ and\ \citenamefont
  {Wegscheider}}]{Charpentier2013}%
  \BibitemOpen
  \bibfield  {author} {\bibinfo {author} {\bibfnamefont {C.}~\bibnamefont
  {Charpentier}}, \bibinfo {author} {\bibfnamefont {S.}~\bibnamefont
  {F{\"{a}}lt}}, \bibinfo {author} {\bibfnamefont {C.}~\bibnamefont {Reichl}},
  \bibinfo {author} {\bibfnamefont {F.}~\bibnamefont {Nichele}}, \bibinfo
  {author} {\bibfnamefont {A.}~\bibnamefont {Nath~Pal}}, \bibinfo {author}
  {\bibfnamefont {P.}~\bibnamefont {Pietsch}}, \bibinfo {author} {\bibfnamefont
  {T.}~\bibnamefont {Ihn}}, \bibinfo {author} {\bibfnamefont {K.}~\bibnamefont
  {Ensslin}}, \ and\ \bibinfo {author} {\bibfnamefont {W.}~\bibnamefont
  {Wegscheider}},\ }\bibfield  {title} {\enquote {\bibinfo {title}
  {{Suppression of bulk conductivity in InAs/GaSb broken gap composite quantum
  wells}},}\ }\href {\doibase 10.1063/1.4821037} {\bibfield  {journal}
  {\bibinfo  {journal} {Appl. Phys. Lett.}\ }\textbf {\bibinfo {volume}
  {103}},\ \bibinfo {pages} {112102} (\bibinfo {year} {2013})}\BibitemShut
  {NoStop}%
\bibitem [{\citenamefont {Akiho}\ \emph {et~al.}(2016)\citenamefont {Akiho},
  \citenamefont {Cou{\"{e}}do}, \citenamefont {Irie}, \citenamefont {Suzuki},
  \citenamefont {Onomitsu},\ and\ \citenamefont {Muraki}}]{Akiho2016}%
  \BibitemOpen
  \bibfield  {author} {\bibinfo {author} {\bibfnamefont {T.}~\bibnamefont
  {Akiho}}, \bibinfo {author} {\bibfnamefont {F.}~\bibnamefont {Cou{\"{e}}do}},
  \bibinfo {author} {\bibfnamefont {H.}~\bibnamefont {Irie}}, \bibinfo {author}
  {\bibfnamefont {K.}~\bibnamefont {Suzuki}}, \bibinfo {author} {\bibfnamefont
  {K.}~\bibnamefont {Onomitsu}}, \ and\ \bibinfo {author} {\bibfnamefont
  {K.}~\bibnamefont {Muraki}},\ }\bibfield  {title} {\enquote {\bibinfo {title}
  {{Engineering quantum spin Hall insulators by strained-layer
  heterostructures}},}\ }\href {\doibase 10.1063/1.4967471} {\bibfield
  {journal} {\bibinfo  {journal} {Appl. Phys. Lett.}\ }\textbf {\bibinfo
  {volume} {109}},\ \bibinfo {pages} {192105} (\bibinfo {year}
  {2016})}\BibitemShut {NoStop}%
\bibitem [{\citenamefont {Du}\ \emph {et~al.}(2017{\natexlab{a}})\citenamefont
  {Du}, \citenamefont {Li}, \citenamefont {Lou}, \citenamefont {Wu},
  \citenamefont {Liu}, \citenamefont {Han}, \citenamefont {Zhang},
  \citenamefont {Sullivan}, \citenamefont {Ikhlassi}, \citenamefont {Chang},\
  and\ \citenamefont {Du}}]{Du2017PRL}%
  \BibitemOpen
  \bibfield  {author} {\bibinfo {author} {\bibfnamefont {L.}~\bibnamefont
  {Du}}, \bibinfo {author} {\bibfnamefont {T.}~\bibnamefont {Li}}, \bibinfo
  {author} {\bibfnamefont {W.}~\bibnamefont {Lou}}, \bibinfo {author}
  {\bibfnamefont {X.}~\bibnamefont {Wu}}, \bibinfo {author} {\bibfnamefont
  {X.}~\bibnamefont {Liu}}, \bibinfo {author} {\bibfnamefont {Z.}~\bibnamefont
  {Han}}, \bibinfo {author} {\bibfnamefont {C.}~\bibnamefont {Zhang}}, \bibinfo
  {author} {\bibfnamefont {G.}~\bibnamefont {Sullivan}}, \bibinfo {author}
  {\bibfnamefont {A.}~\bibnamefont {Ikhlassi}}, \bibinfo {author}
  {\bibfnamefont {K.}~\bibnamefont {Chang}}, \ and\ \bibinfo {author}
  {\bibfnamefont {R.-R.}\ \bibnamefont {Du}},\ }\bibfield  {title} {\enquote
  {\bibinfo {title} {{Tuning Edge States in Strained-Layer InAs/GaInSb Quantum
  Spin Hall Insulators}},}\ }\href {\doibase 10.1103/PhysRevLett.119.056803}
  {\bibfield  {journal} {\bibinfo  {journal} {Phys. Rev. Lett.}\ }\textbf
  {\bibinfo {volume} {119}},\ \bibinfo {pages} {056803} (\bibinfo {year}
  {2017}{\natexlab{a}})}\BibitemShut {NoStop}%
\bibitem [{\citenamefont {Irie}\ \emph {et~al.}(2020)\citenamefont {Irie},
  \citenamefont {Akiho}, \citenamefont {Cou\"edo}, \citenamefont {Ohana},
  \citenamefont {Suzuki}, \citenamefont {Onomitsu},\ and\ \citenamefont
  {Muraki}}]{Irie2020}%
  \BibitemOpen
  \bibfield  {author} {\bibinfo {author} {\bibfnamefont {H.}~\bibnamefont
  {Irie}}, \bibinfo {author} {\bibfnamefont {T.}~\bibnamefont {Akiho}},
  \bibinfo {author} {\bibfnamefont {F.}~\bibnamefont {Cou\"edo}}, \bibinfo
  {author} {\bibfnamefont {R.}~\bibnamefont {Ohana}}, \bibinfo {author}
  {\bibfnamefont {K.}~\bibnamefont {Suzuki}}, \bibinfo {author} {\bibfnamefont
  {K.}~\bibnamefont {Onomitsu}}, \ and\ \bibinfo {author} {\bibfnamefont
  {K.}~\bibnamefont {Muraki}},\ }\bibfield  {title} {\enquote {\bibinfo {title}
  {{Impact of epitaxial strain on the topological-nontopological phase diagram
  and semimetallic behavior of InAs/GaSb composite quantum wells}},}\ }\href
  {\doibase 10.1103/PhysRevB.101.075433} {\bibfield  {journal} {\bibinfo
  {journal} {Phys. Rev. B}\ }\textbf {\bibinfo {volume} {101}},\ \bibinfo
  {pages} {075433} (\bibinfo {year} {2020})}\BibitemShut {NoStop}%
\bibitem [{\citenamefont {Karalic}\ \emph {et~al.}(2016)\citenamefont
  {Karalic}, \citenamefont {Mueller}, \citenamefont {Mittag}, \citenamefont
  {Pakrouski}, \citenamefont {Wu}, \citenamefont {Soluyanov}, \citenamefont
  {Troyer}, \citenamefont {Tschirky}, \citenamefont {Wegscheider},
  \citenamefont {Ensslin},\ and\ \citenamefont {Ihn}}]{Karalic2016}%
  \BibitemOpen
  \bibfield  {author} {\bibinfo {author} {\bibfnamefont {M.}~\bibnamefont
  {Karalic}}, \bibinfo {author} {\bibfnamefont {S.}~\bibnamefont {Mueller}},
  \bibinfo {author} {\bibfnamefont {C.}~\bibnamefont {Mittag}}, \bibinfo
  {author} {\bibfnamefont {K.}~\bibnamefont {Pakrouski}}, \bibinfo {author}
  {\bibfnamefont {Q.S.}\ \bibnamefont {Wu}}, \bibinfo {author} {\bibfnamefont
  {A.~A.}\ \bibnamefont {Soluyanov}}, \bibinfo {author} {\bibfnamefont
  {M.}~\bibnamefont {Troyer}}, \bibinfo {author} {\bibfnamefont
  {T.}~\bibnamefont {Tschirky}}, \bibinfo {author} {\bibfnamefont
  {W.}~\bibnamefont {Wegscheider}}, \bibinfo {author} {\bibfnamefont
  {K.}~\bibnamefont {Ensslin}}, \ and\ \bibinfo {author} {\bibfnamefont
  {T.}~\bibnamefont {Ihn}},\ }\bibfield  {title} {\enquote {\bibinfo {title}
  {{Experimental signatures of the inverted phase in InAs/GaSb coupled quantum
  wells}},}\ }\href {\doibase 10.1103/PhysRevB.94.241402} {\bibfield  {journal}
  {\bibinfo  {journal} {Phys. Rev. B}\ }\textbf {\bibinfo {volume} {94}},\
  \bibinfo {pages} {241402} (\bibinfo {year} {2016})}\BibitemShut {NoStop}%
\bibitem [{Note1()}]{Note1}%
  \BibitemOpen
  \bibinfo {note} {Conventional Arrhenius analysis (i.e., fitting with a single
  exponential function) using high-$T$ data (40 K $\le T \le 70$ K)
  underestimates $\Delta $ by about 8\% compared to $\protect \mathrm {\Delta
  }_{1}$.}\BibitemShut {Stop}%
\bibitem [{\citenamefont {Du}\ \emph {et~al.}(2015)\citenamefont {Du},
  \citenamefont {Knez}, \citenamefont {Sullivan},\ and\ \citenamefont
  {Du}}]{Du2015}%
  \BibitemOpen
  \bibfield  {author} {\bibinfo {author} {\bibfnamefont {L.}~\bibnamefont
  {Du}}, \bibinfo {author} {\bibfnamefont {I.}~\bibnamefont {Knez}}, \bibinfo
  {author} {\bibfnamefont {G.}~\bibnamefont {Sullivan}}, \ and\ \bibinfo
  {author} {\bibfnamefont {R.-R.}\ \bibnamefont {Du}},\ }\bibfield  {title}
  {\enquote {\bibinfo {title} {{Robust Helical Edge Transport in Gated
  InAs/GaSb Bilayers}},}\ }\href {\doibase 10.1103/PhysRevLett.114.096802}
  {\bibfield  {journal} {\bibinfo  {journal} {Phys. Rev. Lett.}\ }\textbf
  {\bibinfo {volume} {114}},\ \bibinfo {pages} {096802} (\bibinfo {year}
  {2015})}\BibitemShut {NoStop}%
\bibitem [{Note2()}]{Note2}%
  \BibitemOpen
  \bibinfo {note} {A similar calculation without self-consistent potential was
  reported in Ref.~\cite {Akiho2016}, where the effect of strain in the InAs
  layer was mistakenly underestimated. Since these two effects tend to
  counteract each other, including them both produces overall similar
  results.}\BibitemShut {Stop}%
\bibitem [{Note3()}]{Note3}%
  \BibitemOpen
  \bibinfo {note} {We used an Arrhenius analysis of high-$T$ data (30 K $\le T
  \le 60$ K) to obtain the $\Delta $ shown in Fig.~\ref {Fig4}(b), as fitting
  the weak temperature dependence near the topological phase transition with
  Eq. (\ref {Eq1}) did not give reliable results.}\BibitemShut {Stop}%
\bibitem [{\citenamefont {Vandenberghe}\ and\ \citenamefont
  {Fischetti}(2017)}]{Vandenberghe2017}%
  \BibitemOpen
  \bibfield  {author} {\bibinfo {author} {\bibfnamefont {W.~G.}\ \bibnamefont
  {Vandenberghe}}\ and\ \bibinfo {author} {\bibfnamefont {M.~V.}\ \bibnamefont
  {Fischetti}},\ }\bibfield  {title} {\enquote {\bibinfo {title} {{Imperfect
  two-dimensional topological insulator field-effect transistors}},}\ }\href
  {\doibase 10.1038/ncomms14184} {\bibfield  {journal} {\bibinfo  {journal}
  {Nat. Commun.}\ }\textbf {\bibinfo {volume} {8}},\ \bibinfo {pages} {14184}
  (\bibinfo {year} {2017})}\BibitemShut {NoStop}%
\bibitem [{\citenamefont {B{\"u}ttner}\ \emph {et~al.}(2011)\citenamefont
  {B{\"u}ttner}, \citenamefont {Liu}, \citenamefont {Tkachov}, \citenamefont
  {Novik}, \citenamefont {Br{\"u}ne}, \citenamefont {Buhmann}, \citenamefont
  {Hankiewicz}, \citenamefont {Recher}, \citenamefont {Trauzettel},
  \citenamefont {Zhang},\ and\ \citenamefont {Molenkamp}}]{Buttner2011}%
  \BibitemOpen
  \bibfield  {author} {\bibinfo {author} {\bibfnamefont {B.}~\bibnamefont
  {B{\"u}ttner}}, \bibinfo {author} {\bibfnamefont {C.~X.}\ \bibnamefont
  {Liu}}, \bibinfo {author} {\bibfnamefont {G.}~\bibnamefont {Tkachov}},
  \bibinfo {author} {\bibfnamefont {E.~G.}\ \bibnamefont {Novik}}, \bibinfo
  {author} {\bibfnamefont {C.}~\bibnamefont {Br{\"u}ne}}, \bibinfo {author}
  {\bibfnamefont {H.}~\bibnamefont {Buhmann}}, \bibinfo {author} {\bibfnamefont
  {E.~M.}\ \bibnamefont {Hankiewicz}}, \bibinfo {author} {\bibfnamefont
  {P.}~\bibnamefont {Recher}}, \bibinfo {author} {\bibfnamefont
  {B.}~\bibnamefont {Trauzettel}}, \bibinfo {author} {\bibfnamefont {S.~C.}\
  \bibnamefont {Zhang}}, \ and\ \bibinfo {author} {\bibfnamefont {L.~W.}\
  \bibnamefont {Molenkamp}},\ }\bibfield  {title} {\enquote {\bibinfo {title}
  {{Single valley Dirac fermions in zero-gap HgTe quantum wells}},}\ }\href
  {\doibase 10.1038/nphys1914} {\bibfield  {journal} {\bibinfo  {journal} {Nat.
  Phys.}\ }\textbf {\bibinfo {volume} {7}},\ \bibinfo {pages} {418} (\bibinfo
  {year} {2011})}\BibitemShut {NoStop}%
\bibitem [{\citenamefont {Mahmoodian}\ and\ \citenamefont
  {Entin}(2020)}]{Mahmoodian2020}%
  \BibitemOpen
  \bibfield  {author} {\bibinfo {author} {\bibfnamefont {M.~M.}\ \bibnamefont
  {Mahmoodian}}\ and\ \bibinfo {author} {\bibfnamefont {M.~V.}\ \bibnamefont
  {Entin}},\ }\bibfield  {title} {\enquote {\bibinfo {title} {{Conductivity of
  a two-dimensional HgTe layer near the critical width: The role of developed
  edge states network and random mixture of $p$- and $n$-domains}},}\ }\href
  {\doibase 10.1103/PhysRevB.101.125415} {\bibfield  {journal} {\bibinfo
  {journal} {Phys. Rev. B}\ }\textbf {\bibinfo {volume} {101}},\ \bibinfo
  {pages} {125415} (\bibinfo {year} {2020})}\BibitemShut {NoStop}%
\bibitem [{\citenamefont {Naveh}\ and\ \citenamefont
  {Laikhtman}(1996)}]{Naveh1996}%
  \BibitemOpen
  \bibfield  {author} {\bibinfo {author} {\bibfnamefont {Y.}~\bibnamefont
  {Naveh}}\ and\ \bibinfo {author} {\bibfnamefont {B.}~\bibnamefont
  {Laikhtman}},\ }\bibfield  {title} {\enquote {\bibinfo {title} {{Excitonic
  Instability and Electric-Field-Induced Phase Transition Towards a
  Two-Dimensional Exciton Condensate}},}\ }\href {\doibase
  10.1103/PhysRevLett.77.900} {\bibfield  {journal} {\bibinfo  {journal} {Phys.
  Rev. Lett.}\ }\textbf {\bibinfo {volume} {77}},\ \bibinfo {pages} {900}
  (\bibinfo {year} {1996})}\BibitemShut {NoStop}%
\bibitem [{\citenamefont {Pikulin}\ \emph {et~al.}(2014)\citenamefont
  {Pikulin}, \citenamefont {Hyart}, \citenamefont {Mi}, \citenamefont
  {Tworzyd\l{}o}, \citenamefont {Wimmer},\ and\ \citenamefont
  {Beenakker}}]{Pikulin2014}%
  \BibitemOpen
  \bibfield  {author} {\bibinfo {author} {\bibfnamefont {D.~I.}\ \bibnamefont
  {Pikulin}}, \bibinfo {author} {\bibfnamefont {T.}~\bibnamefont {Hyart}},
  \bibinfo {author} {\bibfnamefont {Shuo}\ \bibnamefont {Mi}}, \bibinfo
  {author} {\bibfnamefont {J.}~\bibnamefont {Tworzyd\l{}o}}, \bibinfo {author}
  {\bibfnamefont {M.}~\bibnamefont {Wimmer}}, \ and\ \bibinfo {author}
  {\bibfnamefont {C.~W.~J.}\ \bibnamefont {Beenakker}},\ }\bibfield  {title}
  {\enquote {\bibinfo {title} {{Disorder and magnetic-field-induced breakdown
  of helical edge conduction in an inverted electron-hole bilayer}},}\ }\href
  {\doibase 10.1103/PhysRevB.89.161403} {\bibfield  {journal} {\bibinfo
  {journal} {Phys. Rev. B}\ }\textbf {\bibinfo {volume} {89}},\ \bibinfo
  {pages} {161403} (\bibinfo {year} {2014})}\BibitemShut {NoStop}%
\bibitem [{\citenamefont {Du}\ \emph {et~al.}(2017{\natexlab{b}})\citenamefont
  {Du}, \citenamefont {Li}, \citenamefont {Lou}, \citenamefont {Sullivan},
  \citenamefont {Chang}, \citenamefont {Kono},\ and\ \citenamefont
  {Du}}]{Du2017NatCom}%
  \BibitemOpen
  \bibfield  {author} {\bibinfo {author} {\bibfnamefont {L.}~\bibnamefont
  {Du}}, \bibinfo {author} {\bibfnamefont {X.}~\bibnamefont {Li}}, \bibinfo
  {author} {\bibfnamefont {W.}~\bibnamefont {Lou}}, \bibinfo {author}
  {\bibfnamefont {G.}~\bibnamefont {Sullivan}}, \bibinfo {author}
  {\bibfnamefont {K.}~\bibnamefont {Chang}}, \bibinfo {author} {\bibfnamefont
  {J.}~\bibnamefont {Kono}}, \ and\ \bibinfo {author} {\bibfnamefont {R.-R.}\
  \bibnamefont {Du}},\ }\bibfield  {title} {\enquote {\bibinfo {title}
  {{Evidence for a topological excitonic insulator in InAs/GaSb bilayers}},}\
  }\href {\doibase 10.1038/s41467-017-01988-1} {\bibfield  {journal} {\bibinfo
  {journal} {Nat. Commun.}\ }\textbf {\bibinfo {volume} {8}},\ \bibinfo {pages}
  {1971} (\bibinfo {year} {2017}{\natexlab{b}})}\BibitemShut {NoStop}%
\bibitem [{\citenamefont {Leubner}\ \emph {et~al.}(2016)\citenamefont
  {Leubner}, \citenamefont {Lunczer}, \citenamefont {Br\"une}, \citenamefont
  {Buhmann},\ and\ \citenamefont {Molenkamp}}]{Leubner2016}%
  \BibitemOpen
  \bibfield  {author} {\bibinfo {author} {\bibfnamefont {P.}~\bibnamefont
  {Leubner}}, \bibinfo {author} {\bibfnamefont {L.}~\bibnamefont {Lunczer}},
  \bibinfo {author} {\bibfnamefont {C.}~\bibnamefont {Br\"une}}, \bibinfo
  {author} {\bibfnamefont {H.}~\bibnamefont {Buhmann}}, \ and\ \bibinfo
  {author} {\bibfnamefont {L.~W.}\ \bibnamefont {Molenkamp}},\ }\bibfield
  {title} {\enquote {\bibinfo {title} {{Strain Engineering of the Band Gap of
  HgTe Quantum Wells Using Superlattice Virtual Substrates}},}\ }\href
  {\doibase 10.1103/PhysRevLett.117.086403} {\bibfield  {journal} {\bibinfo
  {journal} {Phys. Rev. Lett.}\ }\textbf {\bibinfo {volume} {117}},\ \bibinfo
  {pages} {086403} (\bibinfo {year} {2016})}\BibitemShut {NoStop}%
\bibitem [{\citenamefont {Tang}\ \emph {et~al.}(2017)\citenamefont {Tang},
  \citenamefont {Zhang}, \citenamefont {Wong}, \citenamefont {Pedramrazi},
  \citenamefont {Tsai}, \citenamefont {Jia}, \citenamefont {Moritz},
  \citenamefont {Claassen}, \citenamefont {Ryu}, \citenamefont {Kahn},
  \citenamefont {Jiang}, \citenamefont {Yan}, \citenamefont {Hashimoto},
  \citenamefont {Lu}, \citenamefont {Moore}, \citenamefont {Hwang},
  \citenamefont {Hwang}, \citenamefont {Hussain}, \citenamefont {Chen},
  \citenamefont {Ugeda}, \citenamefont {Liu}, \citenamefont {Xie},
  \citenamefont {Devereaux}, \citenamefont {Crommie}, \citenamefont {Mo},\ and\
  \citenamefont {Shen}}]{Tang2017}%
  \BibitemOpen
  \bibfield  {author} {\bibinfo {author} {\bibfnamefont {S.}~\bibnamefont
  {Tang}}, \bibinfo {author} {\bibfnamefont {C.}~\bibnamefont {Zhang}},
  \bibinfo {author} {\bibfnamefont {D.}~\bibnamefont {Wong}}, \bibinfo {author}
  {\bibfnamefont {Z.}~\bibnamefont {Pedramrazi}}, \bibinfo {author}
  {\bibfnamefont {H.-Z.}\ \bibnamefont {Tsai}}, \bibinfo {author}
  {\bibfnamefont {C.}~\bibnamefont {Jia}}, \bibinfo {author} {\bibfnamefont
  {B.}~\bibnamefont {Moritz}}, \bibinfo {author} {\bibfnamefont
  {M.}~\bibnamefont {Claassen}}, \bibinfo {author} {\bibfnamefont
  {H.}~\bibnamefont {Ryu}}, \bibinfo {author} {\bibfnamefont {S.}~\bibnamefont
  {Kahn}}, \bibinfo {author} {\bibfnamefont {J.}~\bibnamefont {Jiang}},
  \bibinfo {author} {\bibfnamefont {H.}~\bibnamefont {Yan}}, \bibinfo {author}
  {\bibfnamefont {M.}~\bibnamefont {Hashimoto}}, \bibinfo {author}
  {\bibfnamefont {D.}~\bibnamefont {Lu}}, \bibinfo {author} {\bibfnamefont
  {R.~G.}\ \bibnamefont {Moore}}, \bibinfo {author} {\bibfnamefont {C.-C.}\
  \bibnamefont {Hwang}}, \bibinfo {author} {\bibfnamefont {C.}~\bibnamefont
  {Hwang}}, \bibinfo {author} {\bibfnamefont {Z.}~\bibnamefont {Hussain}},
  \bibinfo {author} {\bibfnamefont {Y.}~\bibnamefont {Chen}}, \bibinfo {author}
  {\bibfnamefont {M.~M.}\ \bibnamefont {Ugeda}}, \bibinfo {author}
  {\bibfnamefont {Z.}~\bibnamefont {Liu}}, \bibinfo {author} {\bibfnamefont
  {X.}~\bibnamefont {Xie}}, \bibinfo {author} {\bibfnamefont {T.~P.}\
  \bibnamefont {Devereaux}}, \bibinfo {author} {\bibfnamefont {M.~F.}\
  \bibnamefont {Crommie}}, \bibinfo {author} {\bibfnamefont {S.-K.}\
  \bibnamefont {Mo}}, \ and\ \bibinfo {author} {\bibfnamefont {Z.-X.}\
  \bibnamefont {Shen}},\ }\bibfield  {title} {\enquote {\bibinfo {title}
  {{Quantum spin Hall state in monolayer 1T'-WTe$_{2}$}},}\ }\href {\doibase
  10.1038/nphys4174} {\bibfield  {journal} {\bibinfo  {journal} {Nat. Phys.}\
  }\textbf {\bibinfo {volume} {13}},\ \bibinfo {pages} {683} (\bibinfo {year}
  {2017})}\BibitemShut {NoStop}%
\end{thebibliography}


\end{document}